\documentclass[12pt,preprint]{aastex}
\shortauthors{Basri et al.}  
\shorttitle{Variability in Kepler Target Stars}
\begin{document}

\title{Photometric Variability in Kepler Target Stars: The Sun Among Stars -- A First Look.}

\author{
Gibor Basri\altaffilmark{1}, Lucianne M. Walkowicz\altaffilmark{1}, Natalie Batalha\altaffilmark{2}, Ronald L. Gilliland\altaffilmark{3}, Jon Jenkins\altaffilmark{2}, William J. Borucki\altaffilmark{2}, David Koch\altaffilmark{2}, Doug Caldwell\altaffilmark{2}, Andrea K. Dupree\altaffilmark{4}, David W. Latham\altaffilmark{4}, Soeren Meibom\altaffilmark{4}, Steve Howell\altaffilmark{5}, Tim Brown \altaffilmark{6}
}

\altaffiltext{1}{Astronomy Department, University of California,  Berkeley, CA 94720}
\altaffiltext{2}{NASA Ames Research Center, Moffett Field, CA 94035}
\altaffiltext{3}{Space Telescope Science Institute, Baltimore, MD 21218}
\altaffiltext{4}{Harvard-Smithsonian Center for Astrophysics, Cambridge, MA 02138}
\altaffiltext{5}{National Optical Astronomy Observatory, Tucson, AZ 85719}
\altaffiltext{6}{Las Cumbres Observatory Global Telescope, Goleta, CA 93117}

\begin{abstract}
The \textit{Kepler} mission provides an exciting opportunity to study the lightcurves of stars with unprecedented precision and continuity of coverage. This is the first look at a large sample of stars with photometric data of a quality that has heretofore been only available for our Sun. It provides the first opportunity to compare the irradiance variations of our Sun to a large cohort of stars ranging from vary similar to rather different stellar properties, at a wide variety of ages. Although \textit{Kepler} data is in an early phase of maturity, and we only analyze the first month of coverage, it is sufficient to garner the first meaningful measurements of our Sun's variability in the context of a large cohort of main sequence stars in the solar neighborhood. We find that nearly half of the full sample is more active than the active Sun, although most of them are not more than twice as active. The active fraction is closer to a third for the stars most similar to the Sun, and rises to well more than half for stars cooler than mid K spectral types.

\end{abstract}


\section{Introduction}

Since Galileo and Kelper noted dark spots on the face of the Sun 400 years ago, scientists have been aware that its output is not completely constant. It has only been in the past few decades that precise measurements of the solar irradiance variations have been possible \cite{Frohlich}. These show that the solar output varies by a small amount (a tenth of a percent, or a part in a thousand at most), and that although sunspots really do cause flux to be suppressed, bright faculae counteract this effect in a general way (although not in a one-to-one relationship) so that the integrated irradiance of the Sun is actually larger in the active part of a solar cycle despite clear dips produced by big spot groups. 

It has also been obvious for a long time that some stars are quite variable. There are a plethora of sources of stellar variability, from eclipses to pulsations to accretion phenomena to explosions on various scales, and a host of other possible causes. In the last century we have also been able to identify variations due to starspots, most of which are much larger than sunspots and are found on stars that are considerably more magnetically active \cite{Strassmeier}. Efforts have been made to evaluate what the range and character of brightness variations (both positive and negative) due to magnetic activity on solar-type\footnote{By ``solar-type'' we mean main sequence stars with radiative cores and significant convective envelopes.} stars are, and what stellar characteristics go with them. In general, we can say that more rapidly rotating stars exhibit more variability due to increased magnetism \cite{Pizzolato}. Activity is also related to age, since solar-type stars spin down due to magnetic braking, so that young stars are more active than older ones. It also seems that cooler (relative to the Sun, in the spectral range F-M) stars show high variability more commonly than hotter stars \cite{Hipparcos}, in part because there is greater contrast and temperature sensitivity of heated regions against cooler photospheres (which also allows flares to be more obvious). 

One pressing question has been the place of the Sun in the pantheon of solar-type stars. Suggestions that it might be quieter than typical were made based on a small sample, but this issue has remained unresolved \cite{Lockwood}. This question has gained greater currency as the issue of the effect of solar variability on climate change has come into the public consciousness. It is of interest in any case to know what the range of behavior in solar-type stars is. Proxies for magnetic field (such as CaII H \& K or X-ray emission) have given us a fairly good idea of the general landscape of magnetic activity and the Sun's place in it. Based on emission proxies, Batalha et al. (2002) estimated that about two-thirds of solar-type stars lie in the same range of magnetism as the Sun (minimum to maximum) with one-third more active. This result can arise naturally since by the SunÕs age most of the spindown has already been accomplished, and the Sun is nearly half the age of the Galaxy. It is subject to some imprecision, however, since there is a fair spread of activity at a given stellar rotation rate, and stars exhibit cycles of varying magnitudes. The translation of magnetism into photometric variability is also dependent on spectral type.

Until recently, only variations in fairly active stars have been directly measurable by photometry. The CoRoT satellite has provided the first opportunity for a large-scale survey of photometric variability that approaches the precision needed to resolve variability at the level of the quiet Sun. The first results have been published on a few hundred stars \cite{COROT}; they suggest that almost half of solar-type stars exhibit enough variability that CoRoT can discern a rotation period over timescales up to 80 days or so. While not exact, the activity levels detected seem to be at least that of the active Sun. 

The \textit{Kepler} mission \citep[e.g][]{BoruckiScience} is a wide-field 1-m space telescope whose primary purpose is to detect transits and to discover Earth-size exoplanets. The mission was launched on March 6, 2009. To accomplish its main goal, it obtains photometry of roughly 150,000 stars, most of which are intentionally selected to be solar-type. Although several months of data have been returned, the data reduction process is under refinement. Nonetheless, the first month of data has sufficient precision even in its rough form to undertake an informative analysis of the activity on a very large sample of a wide variety of main sequence stars, and compare their gross variability to that of the Sun.

\section{Observations and Analysis}

The \textit{Kepler} Quarter 1 (Q1) observations took place over $\sim$33.5 days between May 13 and June 15 2009. 156,097 stars were observed at a cadence of $\sim$30 minutes. As the intent of this paper is to determine the intrinsic variability of main sequence stars, we only kept stars that have logg $\ge$ 4 in the \textit{Kepler Input Catalogue} (\textit{KIC}). No cut was made on effective temperature. Of the 156,097 stars observed, 121,432 stars survived this cut, while 24,815 are assigned lower gravity (and therefore presumably subgiants and giants) and 9,849 are unclassified. The \textit{KIC} is known to have some mis-identifications \cite{KochScience}, and some of the unclassified stars are dwarfs, but the errors introduced are thought to be a few percent at most. Since we are primarily interested in variability due to modulation by spots and other manifestations of magnetic activity (at least for stars cooler than 6500K), the sample was further cleaned of obvious transiting/eclipsing systems and contact binaries as described below. The final sample analyzed consisted of 104,376 stars between T$_{eff}$$\sim$3200 and $\sim$19,000 K, with the median T$_{eff}$ around 5600K.   

\subsubsection{Variability Statistics and Periodograms}

The raw flux time series provided by \textit{Kepler} consist of the summed flux in the target pixel aperture, background subtracted and corrected for cosmic ray hits \cite{JenkinsScience}. This does mean that more than one star contributes to it, but the target star is generally the greatest contributor. We first remove a linear slope from the raw flux time series, then a fourth-order polynomial. While this may sometimes remove physical effects, much of these trends are instrumental in nature (as can be inferred from geometrical correlations on the focal plane). These include small shifts of stars on pixels caused by focal changes due to thermal effects from the Sun-angle of the telescope, and changing velocity aberration over the rather large field of view as the velocity vector moves during a quarter. 

The lightcurves are then converted to differential photometry by dividing out the median flux and subtracting unity. The absolute deviations from zero of the rectified lightcurves (boxcar smoothed on a 10-hour timescale for this study) are defined as the ``range'' (half-amplitude) of variability. This will be the primary statistic employed in this paper. A further set of statistics are used to characterize other measures of variability; they are only employed here to filter out non-solar forms of variability. These include metrics to determine the typical timescale of variability: the time separation between points where the differential lightcurve crosses zero, and the time separation between changes in the sign of the slope in the lightcurve. These are easily measured on very large datasets. We calculate the mean, median and maximum for each of our statistics. In addition, we run Horne-Baliunas periodograms \cite{Period} on all of our targets. From the periodograms we determine the significant peaks and then store both the power and location of the individual peaks, as well as the total number of peaks and the power in the most significant peak. This information is then combined with our other statistics to identify subsamples of stars that are likely pulsators and eclipsing systems. This subsample was then examined by eye to ensure that no spot-modulated lightcurves were lost. In future work we will return to the behavior of the full set of stars using all these statistics.

\subsubsection{SOHO Lightcurves}

The SOHO lightcurves provide a straightforward means of comparing the \textit{Kepler} data with that of the Sun as a star. Like \textit{Kepler} photometry, the SOHO bolometric instrument is dominated by a broad white light passband. We also examined the sum of the green and red passbands of the VIRGO instrument, which should also vary like \textit{Kepler} photometry (all these measures vary as a blackbody in each bandpass given the same very small temperature perturbations). We used the 2001 active Sun VIRGO g+r lightcurves, re-binning them to the Kepler cadence ($\sim$ 30 minute sampling) and transforming them into differential fluxes.

The SOHO VIRGO g+r lightcurves were then broken into $\sim$33 day segments of time, one set offset from the other by 16 days, to provide 20 analogs to a \textit{Kepler} lightcurve. The same statistics and periodogram analysis was performed on them. These statistics were then compared with the \textit{Kepler} sample to place the Sun in the context of the \textit{Kepler} observations. 

\section{Results}

\subsection{Variability by brightness and temperature}

\begin{figure}
\begin{center}
\includegraphics[width=6in,bb=0 0 576 432]{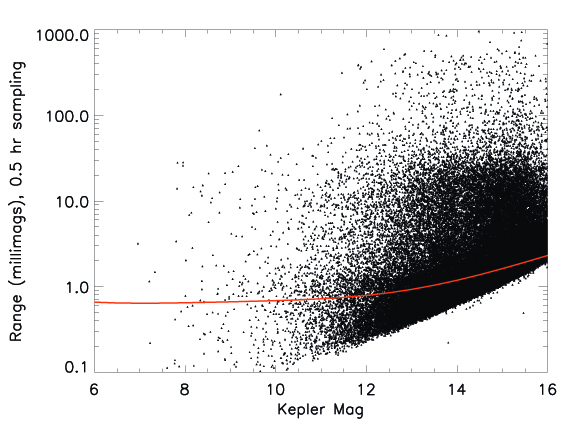}
\end{center}
\caption{Kepler magnitude versus the range of lightcurve modulation. The definition of the range is discussion in Section 2. The red line indicates the locus of the active Sun -- stars lying along this line have levels of activity similar to the active Sun; those that lie below it are quieter, and those above more active. }
\label{kmagvrange}
\end{figure}

The range in variability is our key statistic that best separates active stars from quiet ones -- objects with a large range have large modulation/features, while those with small range are quiet. In Figure \ref{kmagvrange}, we plot the range in millimags as a function of \textit{Kepler} magnitude (Kepmag). The rise of minimum range with fainter magnitude is caused by the smaller stellar signal relative to the fixed instrument noise and increase in shot noise for dimmer stars -- hence, the minimum amount of stellar variability that can be measured is larger for fainter stars. Recall that the range is defined for a smoothed lightcurve; this does not matter for the brighter stars but calms the noise spikes in the fainter stars.

The approximate activity level of the active Sun is shown in this Figure \ref{kmagvrange} as a red line -- lightcurves near this line have modulations similar to the active Sun; those above are more active and those below are less active.  The locus of this line was determined through extensive by-eye comparison between the SOHO and \textit{Kepler} lightcurves. We determined at what value of the range the amplitude of apparently stellar variability for the noisier solar-like stars was consistent with solar-like variability, by looking at many examples with similar range. We will explore a more objective way of doing this later. Across the range of magnitudes, \textit{Kepler} targets which resemble the active Sun were found, although clearly the boundary between these populations is not as sharp as a single line implies. As the noise levels increase, the range has to be bigger to reveal clear variations with the same character as the active Sun. This is why the red line curves up at faint magnitudes. 

\begin{figure}
\begin{center}
\includegraphics[width=6in,bb=0 0 576 432]{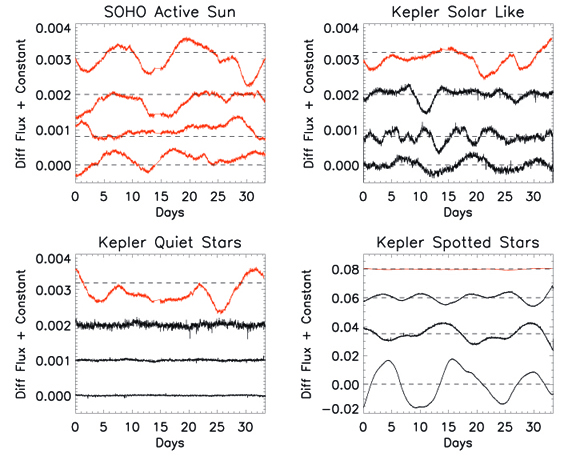}
\end{center}
\caption{Example lightcurves from Kepler and SOHO -- lightcurves are shown offset by various constants for clarity, with their zero levels indicated as dashed lines. Upper left: several segments of SOHO Virgo lightcurves are shown in red. In subsequent panels SOHO lightcurves are shown at the top in red for comparison. Upper right: Kepler lightcurves with activity levels similar to the active Sun. Lower left: quiet Kepler stars. Lower right: several examples of Kepler lightcurves of spotted stars; note that this panel has a markedly different scale than the prior three to show the amplitude of the lightcurve features.}
\label{examples}
\end{figure}

Figure \ref{examples} gives a few examples of lightcurves. The upper left shows a few solar examples from SOHO, while the upper right has a solar example at the top and three Kepler stars we deem comparable below it. The lower left panel shows some quiet Kepler examples (with different noise levels). The lower right panel has a much coarser scale in order to capture the much bigger amplitudes in stars with larger ranges; here the active Sun looks very quiet by comparison at the top. The active fraction is determined by comparing the number of stars below the active Sun (red line) to the numbers of stars above. We find that 46\% of stars in our entire sample are more active than the active Sun. We have also calculated the active fraction of stars as a function of temperature and magnitude bin -- these results are shown In Table 1. Here ``bright'' indicates a Kepler magnitude brighter than 13.5, while ``faint'' indicates stars dimmer than that. 

Figure \ref{teffvrange} shows the range versus effective temperature. The minimum amplitude of variability is larger for cooler stars in the sample, and there is a dearth of quiet cool stars. This is partially, but not entirely due to the fact that cool dwarfs tend to be fainter and noiser. If there were many quiet ones, however, there would be a thick ridge along the noise boundary. The thin vertical gaps in this figure reflect structure in the \textit{KIC}, not real stellar effects. Figure \ref{temppanel} provides an alternative view of activity as a function of magnitude and effective temperature: each of the four panels shows the range for a different bin of effective temperatures as a function of \textit{Kepler} magnitude. Clockwise from upper left it shows the hottest stars ($\sim$21,000 stars, T$_{eff}$ $>$ 6000 K), stars with 6000K $>$ T$_{eff}$ $>$ 5500 K ($\sim$43,000 stars corresponding roughly to late F to mid G dwarfs), stars with 5500K $>$ T$_{eff}$ $>$ 4500K ($\sim$ 35,000 stars from mid G to mid K dwarfs) and T$_{eff}$ $<$ 4500K ($\sim$6500 stars, mid K to mid M dwarfs). The progression of typical range by temperature is evident in these panels, with the hottest stars covering the full space from quiet (low amplitude) to extremely variable (high amplitude), while the majority of the coolest stars (lower right panel) are more variable overall. The vertical features at Kepler magnitudes $\sim$13.9 and $\sim$15.5 in the two hotter panels are artifacts of target selection -- hot stars fainter than selected magnitudes were disfavored to leave more capacity for smaller stars.
 
We also looked at the statistics for stars that are more than twice as variable as the active Sun. Overall this fraction drops to 18\%, showing that a great many of the active stars are not that much more active than our Sun. For the stars in the second (most solar-like) temperature bin, the more active fraction is only 10\%. On the other hand, in the coolest bin we still find 43\% more than twice as active. This result is less sensitive to the noise floor than the straight active Sun fraction, and once again reinforces the impression that the cooler stars are simply more photometrically variable. 

\section{Conclusions and Future Work}

The first thing to remember about these results is that they are rather preliminary. The data reduction pipeline is under development \cite{JenkinsScience}, and much longer time coverage will be available as the mission proceeds. It is also possible to separate out sub-types of lightcurves to a much greater extent than we have done here. This will allow statements not just about the general level of variability, but the separation of magnetic from other sources of variability, and the characterization of the sources of variability. Longer sampling time will help both with confirming short-term behaviors and with capturing variability that occurs on longer timescales. The current sample period is only a bit longer than one solar rotation (although the timescale for features to appear and disappear in the solar lightcurve is comfortably shorter than that). On the other hand, the patching together of data from successive quarterly 90$\deg$ rolls about the optical axis of the spacecraft will require better understanding of the secular trends in the data: which of them are instrumental, which are astrophysical, and how best to remove the instrumental effects.

\begin{figure}
\begin{center}
\includegraphics[width=6in,bb=0 0 576 432]{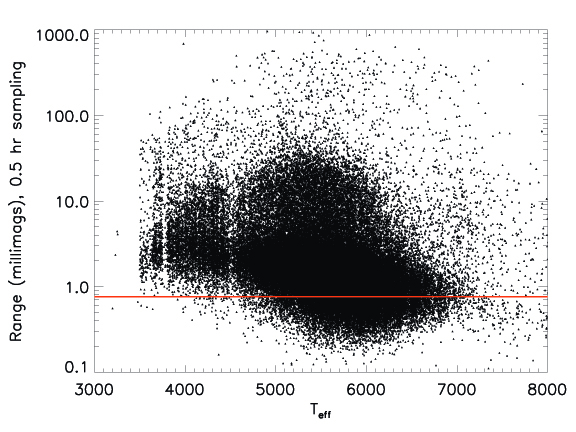}
\end{center}
\caption{Effective temperature from the Kepler Input Catalogue versus the range of lightcurve variability. The red line gives the range value of the active Sun. The dearth of quiet stars at cool temperatures is evident in this figure; this is partly due to the faintness of cool dwarfs and measurement noise.}
\label{teffvrange}
\end{figure}

With these caveats in mind, our preliminary results are fairly clear. As expected, there are a lot of stars more variable than our Sun. In the entire sample nearly half are more active, and this result does not appear to be driven by the noise floor at faint magnitudes. It results from the average of late F to early G stars with active fraction of only a third (these are most comparable to the Sun), late G and K stars whose active fraction is roughly a half, and the even more variable cooler stars. The results for the most Sun-like stars are compatible with the expectations described in Section 1. In the moderately cool stars, the contrast of active areas may increase and/or they have greater magnetic fluxes. The majority of stars cooler than mid-K are much more variable (active fractions $\sim$85\%); this is only partly due to the fact that the noise floor is greater for these fainter stars. Stars hotter than 7000 K are comparably variable to the late F to early G dwarfs but more evenly spread out (Figure \ref{teffvrange}). This is a warning that we are probably looking at more than just magnetic variability, since that is not expected for early F and A stars (except of course the Am and Ap stars). 

\begin{figure}
\begin{center}
\includegraphics[width=6in,bb=0 0 576 432]{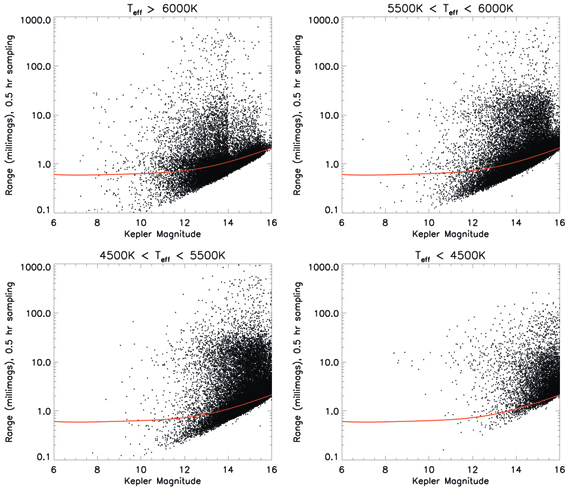}
\end{center}
\caption{Kepler magnitude versus the range of lightcurve variability in different temperature ranges, with the locus of the active Sun shown as red lines. The four panels correspond to different effective temperature bins, clockwise from upper left:  stars hotter than 6000K, stars between 5500 and 6000 K, stars between 4500 and 5500 K, and stars cooler than 4500 K.}
\label{temppanel}
\end{figure}

Figure \ref{teffvrange} appears to show a general trend of increasing variability at lower temperature in the high density region of the plot. This is not primarily an effect of increasing noise for fainter stars, because Figure \ref{temppanel} shows that there are many faint stars in both of the mid-temperature panels (b and c, 5500-6000 K and 4500-5500 K). The high density bulk of stars is also at most three times as variable as the active Sun, and many of them are quieter than the active Sun. There is a thinner distribution of yet more active stars above them, which appears at late F, peaks around early K and declines below mid K. This group reaches a variability level up to 50 times the active Sun. There is a yet sparser set of stars that are even more active than that, which reaches a gentle peak in activity around early K. 

\begin{deluxetable}{lcccc}
\tabletypesize{\scriptsize}
\tablecaption{Fraction of Active Stars}
\tablenum{1}
\tablehead{ \colhead{} & \colhead{T$_{eff}$ $>$ 6000 K} & \colhead{5500 K $<$ T$_{eff}$ $<$ 6000 K} & \colhead{4500 K $<$ T$_{eff}$ $<$ 5500 K} & \colhead{T$_{eff}$ $<$ 4500 K} } 
\startdata
\hline
all stars            & 0.36 & 0.37 & 0.57 & 0.87 \\
N$_{stars}$, total    & 21023 & 42832 & 33288 &6522 \\
bright stars & 0.41 & 0.31 & 0.46 & 0.84 \\
N$_{stars}$, bright  & 8747 & 6164 & 2613 & 369 \\ 
faint stars & 0.33 & 0.38 & 0.57    & 0.88 \\
N$_{stars}$, faint & 12276 & 36668 & 30675 & 6153 \\
\enddata
Note: The dividing line between bright and faint is taken at Kepmag=13.5.
\end{deluxetable}

One goal of our future work is to determine rotation periods for as many of the \textit{Kepler} stars as possible. Since rotation is intimately linked to the magnetic activity of stars, spots provide a natural way to trace the stellar rotation period (except for the quietest stars). It is clear that there are tens of thousands of \textit{Kepler} targets which exhibit sufficient variability to allow determination of their rotation period. As improvements in the data occur and the length of coverage keeps increasing, even more stars will yield their rotations. It will be possible to define a rotation-activity relation purely from the  \textit{Kepler} data, since the amplitude of variability may be related to activity for spotted stars. Of course, there are many potential complications. For example, polar or widely distributed spots might not produce a recognizable signal, bright faculae produce a somewhat opposite effect, and the solar analogy might break down to various degrees. In the cooler stars, we will also be able to study the flaring fraction. Beyond determining periods, we will be able to model the spot coverage fraction, average lifetime and growth and decay rates, and track differential rotation among different spots on the same star in many cases. It is more difficult to constrain the stellar inclination, and polar spots will only show up as slow variations. Over the few years of the mission, it may be possible in some cases to discern the stellar activity cycle. The potential of the \textit{Kepler} mission to contribute to the understanding of solar-type stars via photometric variability due to magnetic activity is extremely promising.

\acknowledgements
The authors wish to thank the entire Kepler mission team, including the engineers and managers who were so pivotal in the ultimate success of the mission. LW is grateful for the support of the \textit{Kepler Fellowship for the Study of Planet-Bearing Stars}. Funding for this Discovery mission is provided by NASA's Science Mission Directorate.


\end{document}